\documentstyle[preprint,aps]{revtex}
\begin{document}
\draft
\title {AN INVESTIGATION OF THE PHASE TRANSITION IN
        THE EXTENDED BAG MODEL}
 
 \author{V. Gogohia, B. Luk\'acs and M. Priszny\'ak}
 
\address{RMKI, Department of Theoretical Physics,
         Central Research Institute for Physics,  \\
         H-1525,  Budapest 114,  P. O. B.  49, Hungary }
 
\maketitle
 
\begin{abstract}
By proposing an Ansatz for the pressure (measured in terms
of the bag constant) of the hadronic gas in equilibrium, we have
formulated a rather simple phenomenological model (the extended
bag model) which allows one
to analytically investigate the bulk thermodynamic propertities in
the vicinity of the phase transition. This makes it also possible
to take into account the nonperturbative vacuum effects from both
sides of the equilibrium condition. As an example of our approach,
we have calculated the crossover (critical or transition)
temperature $T_c$ and the critical chemical potential $\mu_c$ (as
functions of the bag constant) from the noninteracting
quark-gluon plasma state equation. Our results for
$T_c(N_f=0)=241.5 \ MeV$ and $T_c(N_f=2)=160.6 \ MeV$ are
in good agreement with recent lattice data. The extensive
densities such as the entropy density, specific heat, etc are
calculated as well.
A general scheme how to calculate the latent heat, the critical
energy density, etc within the extended bag model is also described
and it is applied to the two models of the hadronic gas phase.
\end{abstract}


\vfill
\eject

   I. The quark-gluon plasma (QGP) phase is a necessary
step in the evolution of the Exited Matter from Big Bang to the
present days. Apparently, the only way to study this phase
of the expansion of the Universe is nuclear (heavy ion) collisions
at high energies which makes it possible "to recreate conditions
akin to the first moments of the Early Universe, the Big Bang, in
the laboratory" [1]. Because of the confinement phenomenon,
the nonperturbative vacuum structure must play a very
important role in the transition from QGP to the formation of the
hadronic particles (i. e., hadronization) and vice versa. As it
was underlined in our papers [2], any correct model of the
nonperturbative effects such as quark confinement or dynamical
chiral symmetry breaking (DCSB) becomes a model of the true QCD
ground state (i. e., the nonperturbative vacuum) and the other way
around. Thus the difference between the perturbative (always
normalizable to zero) and the nonperturbative vacua appears to
be necessarily nonzero and finite to describe the above mentioned
nonperturbative phenomena at zero temperature. The existence of
the finite vacuum energy per unit volume - the bag constant [3, 4]
 - becomes important for a realistic calculation of the transition
data from the hadronic gas (HG) to QGP phases at nonzero
temperature as well.
 
There are two main approaches to investigate QGP,
namely the resummed finite-temperature perturbation theory
(effective field theory method) [5, 6] and the lattice one [7].
The former breaks down after the fifth
order in the QCD coupling constant $g$ because of the severe
infrared divergences in the Braaten-Pisarski-Kapusta (BPK) series
in powers of $g^{m+2n} \ln^n g$ [5, 6, 8], but it smoothly
incorporates the case of nonzero chemical potential(s).
The latter is a powerful nonperturbative tool to calculate
equations of state for both phases. However, up to now, there
are no realistic lattice data available for nonzero chemical
potential(s) (for problems to introduce it on the
lattice see, for example, recent paper [9]).
 
 As was emphasized in Ref. [10], at present the phase transition
at finite baryon chemical potential(s) can be only studied
within the phenomenological models. The most popular among them
are, of course, the bag-type models [11], which differ from each
other by modelling the equation of state of the hadronic phase
[12, 13].
Within the framework of the bag-type models, a phase transition
between the HG and QGP phases is constracted
via the Gibbs criteria for a phase equilibrium. So the phase
transition is necessary of the first order. This means that the
thermodynamic quantities of interest are discontinuous
across the critical curve. Our main goal here is to propose
a new bag-type model, the so-called extended bag model. It
complements the standard bag model by an Ansatz for the
pressure (measured in terms of the bag constant) which allows one
to correctly
take into account the nonperturbative vacuum effects from both
sides of the equilibrium condition. We have calculated the critical
chemical potential $\mu_c$ and, especially, the crossover
(critical or transition) temperature $T_c$, at which deconfinement
phase transition can occur and chiral symmetry is
restored, in terms of the bag constant. Our numerical results
for $T_c$ are in good agreement with recent lattice data (see
below). The extended bag model makes it also possible to
analytically investigate the bulk thermodynamic propertities in
the vicinity of the phase transition.
 
II. In the bag-type models the
equation of state of the QGP phase is usually approximated
by the thermal perturbation theory as the ideal (noninteracting)
gas consisting of gluons and massless quarks. It
determines the dependence of the QGP thermodynamical quantities
such as energy density $\epsilon$ and pressure $P$
on the thermodynamical variables, temperature $T$ and quark
chemical potentials $\mu_f$. There exist excellent reviews on the physics
of the QGP (see, for example, Refs. [14, 15]), as well as on the
phase transitions in it [16]. The pressure (i. e., the
thermodynamic
potential $\Omega$, apart from the sign) for the noninteracting
QGP is given as follows [15]
\begin{equation}
P = {1 \over 3} f_{SB} T^4 + {N_f \over 2} \mu_f^2 T^2
+ {N_f \over 4 \pi^2} \mu_f^4 - B,
\end{equation}
where $B$ is the bag constant (see below),
while $N_f$ is the number
of different quark flavours. In what follows we will consider the
values $N_f = 0, 1, 2$ since the inclusion of the strange $(s)$ quark
requires a special treatment [17].
The value $N_f = 0$ describes the case of the pure gluon plasma.
Note also that the state equation (1) is derived by
neglecting quark current masses.
The constant $f_{SB}$, entering the equation of state (1),
\begin{equation}
 f_{SB} \equiv f_{SB} (N_f) =
{\pi^2 \over 5} \Bigl({8 \over 3} + {7 \over 4} N_f \Bigr)
\end{equation}
is the Stefan-Boltzmann (SB) constant, which determines
the ideal (noninteracting gluons and massless quarks) gas limit.
Obviously, for $N_f=0$ it equals to the standard SB
constant of the ideal gluon gas. It is well known that the
energy density $\epsilon$ of
the noninteracting QGP can be obtained from the thermodynamic
pothential (1) as follows
\begin{equation}
\epsilon = 3 P + 4 B,
\end{equation}
i. e. the bag pressure $B$ determines the deviation from the
ideal gas relation. Let us make a few remarks in advance. Our
calculations of $T_c$ and $\mu_c$
for the noninteracting QGP are not based on the bag model state
equation (3). The constraint, determining
the phase transition, will be obtained with the help of an Ansatz
which is beyond the standard bag model and it is general (see
below, part III). We will use
numerical values of the bag constant which was obtained from a
completely different source, namely it was calculated in the
zero modes enhancement (ZME) model of the true QCD vacuum
on account of the instanton contributions as well.
(see below, part IV).
 
III. The Gibbs conditions for the phase equilibrium between
the HG and QGP phases at $T=T_c$ are formulated as
follows [11]
\begin{equation}
P_h = P_q = P_c; \quad T_h = T_q = T_c; \quad
3 \mu_f = \mu_c,
\end{equation}
where subscripts h, q and c refer to HG, QGP phases and
transition (critical or crossover) region, respectively. At the
same time, the difference between $\epsilon_q - \epsilon_h$ at
$T=T_c$ can remain finite (nonzero) and determines
the latent heat (LH), $\epsilon_{LH}$.
 
  Let us formulate our primary assumption (Ansatz) now.
The state equation for
the hadronic phase (the left hand side of the equilibrium
condition (4)) is strongly model dependent [11-13]. However,
in any model the pressure $P_h$ at any values of $T$ and $\mu_h$,
in particular at $T=T_c$ and $\mu_h = \mu_c$ , can be measured
in terms of the above mentioned bag constant, i.e. let us put
\begin{equation}
 P_h (T_c, \mu_c) = { b_h \over a_h + N_f} B,
\end{equation}
where the so-called parametric functions $b_h \equiv b_h (T_c,
\mu_c)$
and $a_h \equiv a_h (T_c, \mu_c)$ describe the details of the HG
phase at the phase boundary. Evidently, they may only depend
on the set of independent dimensionless variables (by
definition) which characterize the HG phase. For example,
\begin{eqnarray}
a_h \equiv a_h (T_c, \mu_c) = a_h (x_c, t_c, y_c, z_c), \nonumber\\
b_h \equiv b_h (T_c, \mu_c) = b_h (x_c, t_c, y_c, z_c),
\end{eqnarray}
where
\begin{equation}
x_c = {\mu_c \over T_c}, \quad  t_c = {\tilde{B}^{1/4} \over T_c},
\quad y_c = {\mu_c \over m}, \quad z_c = \mu_c R_0,
\end{equation}
and $m$ denotes the hadron mass while $R_0$ denotes radius of the
nucleon, so it allows one to take into account finite size
effects due to hard core repulsion between nucleons (extended
volume corrections) [18]. These variables are independent and all
other possible dimensionless variables are obtained by combination
of these, for example, $R_0 T_c = z_c / x_c, \quad m / T_c =x_c /
y_c, \quad \mu_c / \tilde{B}^{1/4} = x_c / t_c, etc$. Also the set
of independent dimensionless variables at the phase boundary (7)
may be extended in order to treat the HG phase in more
sophisticated way. However, in any case the parametric
functions should be symmetric, i. e.
$a_h (T_c, \mu_c)= a_h (- T_c, - \mu_c)$ and
$b_h (T_c, \mu_c)= b_h (- T_c, - \mu_c)$.
 
From
Eqs. (1) and (4-5) at $T=T_c$ and $\mu_f = \mu_c / 3$, one obtains
\begin{equation}
f_{SB}(N_f) T^4_c + {N_f \over 6} \mu_c^2 T^2_c
+ {N_f \over 108 \pi^2} \mu_c^4 = {3 \over a_h + N_f} \tilde{B},
\end{equation}
where we introduced a new "physical" (effective) bag constant
as follows
\begin{equation}
\tilde{B} =  (b_h + a_h + N_f) B,
\end{equation}
and it linearly depends on $N_f$ as it should be at log-loop level
(see below, part IV).
 In connection with our Ansatz (5) a few remarks are in order.
Its alternative parametrization with respect to $N_f$, namely
$P_h (T_c, \mu_c) = (b_h^{'} / a_h^{'} N_f + 1)B$ leads to the
effective bag constant as $\tilde{B} =(b_h^{'} + a_h^{'}N_f +1)B$.
The parameteric functions $b_h$ and $a_h$, as well as $b_h^{'}$
and $a_h^{'}$, as was mentioned above, may, in
principle, arbitrary depend on dimensionless variables (7).
If, for example, $a_h^{'}$ vanishes at
$\mu_c =0$ then the linear dependence of $\tilde{B}$ on $N_f$ will
be spoiled. In other words, the choosen parametrization guarantees
the linear dependence of $\tilde{B}$ on $N_f$ and the
alternative one does not.
 
From now on  $T_c$
and $\mu_c$ will be calculated in terms of $\tilde{B}$ and not
that of old $B$, i. e. a definite numerical value
will be assigned to $\tilde{B}$.
So we consider $\tilde{B}$ as the physical bag constant,
while $B$ as unphysical "bare" one.
The bag constant is a universal one and it represents the complex
nonperturbative structure of the QCD true vacuum. Thus the
proposed Ansatz, allows one to take into account nonperturbative
vacuum effects (parametrized in terms of $\tilde{B}$)
from both sides of the equilibrium condition (4).
However, it still remains dependent on the arbitrary
parameter $a_h$. In order to eliminate this dependence, let us
normalize the thermodynamic potential at the phase boundary in Eq.
(8) to the standard SB constant (2). This yields
\begin{equation}
\tilde{f}_{SB} (N_f) = (N_f + a_h) f_{SB} (N_f) = f_{SB}(0) \quad
at \quad N_f = 0,
\end{equation}
and one immediately arrives at $a_h \equiv a_h (T_c, \mu_c) = 1$.
This is our normalization condition and it leads to good numerical
results for $T_c$ and $\mu_c$ (see below).
So the general constraint (8) finally becomes
uniquely determined, namely
\begin{equation}
f_{SB} T^4_c + {N_f \over 6} \mu_c^2 T^2_c
+ {N_f \over 108 \pi^2} \mu_c^4 = {3 \over N_f + 1} \tilde{B},
\end{equation}
and consequently allows one to investigate the bulk
thermodynamic quantites in the vicinity of a critical point.
Let us emphasize the important observation that the numerical
values of $T_c$ and $\mu_c$ calculated from the constraint (11)
do not depend on how one approximates the equation of state of the
hadronic phase. In contrast to the standard bag-type models
(differed from each other by modelling the hadronic phase), in the
extended bag model their values depend only on $\tilde{B}$ which
incorporates nonperturbative vacuum effects from both sides of the
Gibbs equilibrium condition (4) as it has been already emphasized
above.
 
 The QGP energy density and pressure, however, remain undetermined
with our Ansatz (5) at this stage. In terms of $\tilde{B}$ they become
\begin{equation}
P_h (T_c) = P_q (T_c) ={b_h \over (N_f + 1)[(N_f + 1)+ b_h ]}
\tilde{B}
\end{equation}
and because of Eq. (3),
\begin{equation}
\epsilon_q (T_c) \equiv \epsilon_c = {3 b_h + 4 (N_f + 1) \over
(N_f +1) [(N_f +1) + b_h]} \tilde{B}.
\end{equation}
As mentioned above, the
unknown $b_h$ reflects the fact that the hadronic phase
state equation is strongly model dependent. The unknown $b_h$
is the price paied to determine the above mentioned physical
quantities with our Ansatz. Precisely for this reason, the
bag model state equation (3) plays no role in our numerical
investigation of the phase transition with the constraint (11)
from which $T_c$, as well as $\mu_c$, can be derived. Below (see
part VI) a general scheme, how to calculate $b_h$ within our
model, will be developed. Concluding this part, let us note
that our model cetainly requires the coexistence regime between
QGP and HG phases since $\epsilon_q (T_c)$ as given by Eq. (13)
explicitly
depends on $b_h$ which describes details of the HG phase at
$T=T_c$.
 
IV. Let us discuss the possible values of the bag constant itself
now. The bag constant is the difference between the
energy density of the perturbative and the nonperturbative
QCD vacua (at zero temperature). The former one can always be
normalized to zero, so the bag constant is defined as
$\tilde{B} = - \epsilon$,
where $\epsilon$ is the energy density of the nonperturbative
vacuum. This is always negative, consequently the bag constant
is always positive.
  Let us start from the so-called standard value.
In the random instanton
liquid model (RILM) [19] of the QCD vacuum, for a dilute
ensemble, one has
$\epsilon_I = -(1/4) (11 - {2 \over 3} N_f) \times 1.0 \ fm^{-4}$.
 Then the standard value of the bag constant is
\begin{equation}
\tilde{B}_{st} \equiv \tilde{B}_I = - \epsilon_I = (0.00417 - N_f 0.00025)
\ GeV^4.
\end{equation}
Thus one can conclude in that the standard value of the bag
constant is determined by the instanton component of the
nonperturbative QCD vacuum only. Note that for $N_f=3$ it coincides
with the estimate of the QCD sum rules approach on account of the
phenomenological value of the gluon condensate [20]. From (14) it
also
follows that the difference between values of the standard bag constant
for different values of $N_f$ is rather small but not negligible.
But the main problem with the value of the bag constant, as given
by expression (14), is, of course, its wrong dependence on $N_f$.
It decreases with increasing $N_f$ that defies a physical
interpretation of the bag constant as the energy per unit volume.
That is why the value of the bag constant at the expense of the
instanton contributions only is at least not complete.
 
   Instantons are classical solutions to the dynamical equations of
motion of nonabelian gluon fields which contribute to the
nonperturbative
vacuum energy density. Therefore they are unable to explain
confinement phenomenon, which, no doubt, is a quantum
nonperturbative effect. The QCD vacuum has a much more remarkable
(richer) topological structure than instantons alone can provide.
The dynamical mechanisms
of such nonperturbative effects as quark confinement and DCSB are
closely related to the complicated topological large scale
structure of the QCD true vacuum. Assuming that the low-frequency
modes of the Yang-Mills fields can be enhanced due to the possible
nonperturbative IR divergences in the true QCD vacuum [21], we
have recently proposed the zero modes enhancement
(ZME) model of the true QCD vacuum [2]. This is based on
the solution to the Schwinger-Dyson (SD) equation for the quark
propagator in the infrared domain. We have shown that this model
reveals several desirable and promising features. A single quark
(heavy or light) is always off mass-shell, i.e. the quark
propagator has no poles. It also implies DCSB at the fundamental
quark level, i.e. a chiral symmetry preserving solution is
forbidden and a chiral symmetry violating solution is required.
We have calculated contributions to the vacuum energy density
at log-loop level [2], coming from the confining
quarks with dynamically generated masses, $\epsilon_q$, and
of the nonperturbative gluons, $\epsilon_g$,  due to the
enhancement of zero modes.
These contributions are almost the same and the sum is
$\epsilon = \epsilon_q + \epsilon_g = - 0.0015 \ GeV^4 -
0.0016 \ GeV^4 = - 0.0031 \ GeV^4$. Thus the value of the bag
constant, as given by the ZME model, is
\begin{equation}
\tilde{B}_{ZME} = -( \epsilon_g +
N_f \epsilon_q) = (0.0016 + N_f 0.0015 ) \ GeV^4,
\end{equation}
where we introduced the dependence on $N_f$ since $\epsilon_q$
itself gives a single confining quark contribution to the vacuum
energy density. However, neither contributions (15) nor (14) are
complete. In the above mentioned papers [2], it was already
explained in detail why the instanton-type fluctuations are needed
for the ZME model. It has been proposed there
to add $\tilde{B}_I$, given by Eq. (14), to ZME model value
(15) in order to get a more realistic value of the bag constant.
Thus one obtains
\begin{equation}
\tilde{B} = \tilde{B}_I + \tilde{B}_{ZME} = - \epsilon_t = -( \epsilon_I +
\epsilon_g + N_f \epsilon_q) = (0.00577 + N_f 0.00125 ) \ GeV^4.
\end{equation}
Numerically our values are as follows
\begin{eqnarray}
\tilde{B} (N_f=0) &\simeq& 0.006 \ GeV^4
\simeq (278 \ MeV)^4 \simeq 0.78 \ GeV/fm^3, \nonumber\\
\tilde{B} (N_f=1) &\simeq& 0.007 \ GeV^4
\simeq (290 \ MeV)^4 \simeq 0.91 \ GeV/fm^3, \nonumber\\
\tilde{B} (N_f=2) &\simeq& 0.008 \ GeV^4
\simeq (300 \ MeV)^4 \simeq 1.04 \ GeV/fm^3.
\end{eqnarray}
We will use these values of the bag constant. Evidently, our
value (16) overestimate the MIT bag model value of the bag
constant [9] at least by one order of magnitude.
 
 All values of the bag constant below the so-called standard value
(14) as well as the standard value itself should be ruled out
since it does not account for all
components of the true QCD vacuum, as we pointed out above
(see also Refs. [2]).
There exist already phenomenological estimates [22] as well as
lattice calculations [23] preffering a bigger-than-standard value
of the bag constant. The above described components produce the
main (leading) contribution to the vacuum energy density and
consequently to the bag constant. The next-to-leading
contributions (as given by the effective potential for
composite operators at two-loop level [24]) are $h^2$-order, where
$h$ is the Plank constant. Thus they are suppressed at least by
one order of magnitude in comparison with our values (16).
It has been noticed in Ref.
[1] that nobody knows yet how big the bag constant might be,
but generally it is thought that it is about $ 1 \ GeV/fm^3$. The
proposed value (17) for the most realistic case of the two
thermal quark species ($N_f=2$), which
is more or less realized in heavy ion collisions at high
energies, is in agreement with this expectation.
 
  V. Phase diagram $(T_c, \mu_c)$ for the physically relevant
case $N_f= 2$, as determined from the critical curve (11), is
shown in Fig. 1. For end points of this curve $(T_c, 0)$ and
$(0, \mu_c)$ our data are
\begin{eqnarray}
T_c (N_f=0) = 241.5 \ MeV, \nonumber\\
T_c (N_f=1) = 186.8 \ MeV,  \nonumber\\
T_c (N_f=2) = 160.6 \ MeV,
\end{eqnarray}
and
\begin{eqnarray}
\mu_c (N_f=1) = 1833.8 \ MeV, \nonumber\\
\mu_c (N_f=2) = 1441.4 \ MeV,
\end{eqnarray}
respectively. Thus our value of $T_c$ for the physically relevant
case of the two quark species $N_f=2$ at $\mu_c=0$ is in fair
agreement with its best numerical estimate which comes from the
observed
spectrum of low-$p_T$ secondaries in high-energy hadron-hadron
collisions, according to the analysis of Hagedorn (see Ref. [25]
and references therein).
 
  Let us compare our results for $T_c$ with the
finite temperature lattice QCD simulations with two light
staggered (Kogut-Susskind) quarks represented in Ref. [26].
This can be done by means of our data shown in Eq. (18) at
$\mu_c =0$ only since, to our best
knowledge, until now there are no realistic lattice data available
for nonzero chemical potentials.
The lattice result is
$T_c (N_f=2) = 155(9) \ MeV$
with a systematic uncertanty of about $15\%$, so the agreement
with our value (18) is rather good.
From our data (18) it also follows
that the critical temperature $T_c$ for the pure $SU(3)$ gauge
theory ($N_f=0$) is much higher than for the full QCD with two flavors
($N_f=2$). This is in agreement with lattice
calculations [26, 27], as well as with arguments of a simple
percolation model [27].
  A rather small discrepancy between our value of the critical
temperature for the pure $SU(3)$ gauge theory displayed in
Eqs. (18) for $N_f=0$ and recent lattice calculation
$T_{latt}(N_f=0) \simeq 260 \ MeV$ [28] can be explained as
follows. From the constraint (11) at $\mu_c=0$ one obtains
\begin{equation}
T_c = \Bigl( {3 \over (N_f +1) f_{SB}(N_f)} \Bigr)^{1/4}
\tilde{B}^{1/4}.
\end{equation}
In lattice approach $T_c$ is usually calculated from the string
tension in accordance with $ T_c / \sqrt \sigma = 0.629(3)$ [29].
This ratio depends on $N_f$, i. e. it is smaller for $N_f= 2$ than
for $N_f= 0$, while the string tension itself does not depend on
$N_f$ [27]. Using its standard value $\sqrt \sigma = 420 \ MeV$,
one obtains the above mentioned estimate. In a similar way, let us
evaluate the coefficient in our expression (20) at $N_f= 0$,
while for the bag constant let us abstract from its dependence
on $N_f$ and use its value for the physically relevant (at
nonzero temperature) case of the two light quark species, i. e.
$\tilde{B}^{1/4} = 300 \ MeV$ (see Eqs. (17)). One immediately
obtains $T_c = 260.6 \ MeV$ in fair agreement with the above
mentioned recent lattice calculations [28, 29].
However, the numerical results of quenched (pure
gauge theory for staggered quarks) finite-temperature lattice QCD
should be (perhaps slightly) reconsidered in the light of the
"hard chiral logarithms" problem (see recent review [30] and
references therein).
 
VI. One of the attractive features of our approach is that the
thermodynamic quantities which are defined as derivatives
of the thermodynamic potential $\Omega=-P$
(extensive densities, such as the entropy density, specific
heat, etc) are uniquely determined at the phase boundary.
However, let us begin with introducing
the metric which shows the existence of nontrivial fluctuations in
QGP [31, 32]. This is done in two steps. First, the specific heat
matrix is defined as follows
\[ \Delta = \begin{array}{|lr|}
\partial^2 P / \partial T^2 & \partial^2 P / \partial T \partial
\mu_f \\ \partial^2 P / \partial T \partial \mu_f & \partial^2 P
/ \partial \mu_f^2
\end{array}                \]
The corresponding entropic potential $P/T$ possesses a statistical meaning
[31, 32]. The matrix of its second derivatives
$g_{ik} = \partial^2 (P/T) / \partial x^i \partial x^k$
with respect to the canonical coordinates $x^i \equiv (1 / T,
- \mu_i/T)$ defines the avarage scales of the fluctuations by
$<g_{ik} \delta x^i \delta x^k> = 1$. There is an intimate connection
between $g_{ik}$ and the specific heat determinant $\Delta_{ik}$
which shows that if the former is positive definite then the latter is also
positive definite and the other way around [31, 32]. Then the
fluctuations
remain finite. For the noninteracting medium the specific heat
determinant is always positive definite. Using Eqs. (1-2), one
obtains
$\Delta(T, \mu_f) = 4 f_{SB} T^4 + {3 \over \pi^2} N^2_f \mu_f^4
+ (6.4 +1.2 N_f) N_f T^2 \mu_f^2$.
At the transition phase it finally becomes (on account of Eq.
(11)) $\Delta(T_c, \mu_c) = 4 N_f \Bigl[{3 \over N_f +1} \tilde{B}
+ (0.177 - 0.133 N_f) T^2_c \mu_c^2 \Bigr]$.
For $N_f =0, 1$ it is automatically positive definite and, using
our numerical results for $N_f =2$, it is also positive. Thus
there are no nontrivial fluctuations in the noninteracting QGP
in the vicinity of the phase transition as it should be indeed.
 
  The entropy density is defined as
$s =  \Bigl( {\partial P \over \partial T} \Bigr)_{\mu_f}$,
so, on account of Eqs. (1-2), it is,
$s = {4 \over 3} f_{SB} T^3 + N_f \mu_f^2 T$.
At the phase boundary it finally becomes
\begin{equation}
s (T_c, \mu_c) = {4 \over 3} f_{SB} T^3_c + {N_f \over 9} \mu_c^2
T_c.
\end{equation}
Its behaviour along the critical curve (11) is shown in Figs. 2.
 
The specific heat is defined as follows
$c = T \Bigl( {\partial^2 P \over \partial T^2} \Bigr)_{\mu_f}$,
and again using Eqs. (1-2), one obtains
$c = 4 f_{SB} T^3 + N_f \mu_f^2 T$.
At the phase transition it finally becomes
\begin{equation}
c (T_c, \mu_c) = 4 f_{SB} T_c^3 + {N_f \over 9} \mu_c^2 T_c.
\end{equation}
Its behaviour along the critical curve (11) is shown in Figs. 3.
 
The net quark number density is
$n_f =  \Bigl( {\partial P \over \partial \mu_f} \Bigr)_T$.
Again using Eqs. (1-2), one obtains
$n_f = N_f \mu_f (T^2 + {1 \over \pi^2} \mu_f^2)$.
Since the baryon (B) number
of a quark is $1/3$, the net baryon number density
$n_B = {1 \over 3} n_f$ at the phase transition finally becomes
\begin{equation}
n_B (T_c, \mu_c) = {N_f \mu_c  \over 9} (T^2_c + {1 \over
9 \pi^2} \mu_c^2).
\end{equation}
Its behaviour along the critical curve (11) is shown in Figs. 4.
 
 One of the important observables measured in heavy ion collisions
 is the specific entropy per baryon, $s/n_B$ (see, Refs. [10, 33]
and references
therein). Its behaviour across and along the phase transition may
shed light on the strangeness production as a signal of QGP
formation in heavy ion collisions. At the phase transition
it is determined as follows
\begin{equation}
{s \over n_B } (T_c, \mu_c) = {s (T_c, \mu_c) \over n_B (T_c, \mu_c)},
\end{equation}
where $s (T_c, \mu_c)$ and $n_B (T_c, \mu_c)$ are given by
Eqs. (21) and (23), respectively. Its numerical values can be
easily obtained from curves in Figs. 5.
In Fig. 1, the path of constant
$s /n_B \simeq 50$, which is expected from the QGP fireball, is
additionally shown. From Fig. 1 it also follows that for the QGP
fireball temperature $T \simeq 215 \pm 10 \ MeV$ the baryochemical
potential is $\mu_B \simeq 340 \pm 20 \ MeV$ as it should be
indeed [33].
 
Let us now compute the numerical values of the entropy
density $s$ and specific heat $c$ at $\mu_c =0$ since the
dependence on temperature is more important than the dependence on
chemical potential. From Eq. (21) and constraint equation (11) one
obtains
\begin{equation}
s_c = {4 \over 3} f_{SB}^{1/4} \Bigl( {3 \over N_f +1} \tilde{B}
\Bigr)^{3/4}.
\end{equation}
Numerically this gives
\begin{eqnarray}
s_c (N_f=0) = 0.0992 \ GeV^3, \nonumber\\
s_c (N_f=1) = 0.0751 \ GeV^3,  \nonumber\\
s_c (N_f=2) = 0.0666 \ GeV^3.
\end{eqnarray}
The numerical values for the specific heat at the phase boundary
$c_c$ may be obtained from Eq. (26) in accordance with the
relation $c_c = 3 s_c$, which comes from Eqs. (21) and (22).
 
VII. Here let us develop a general
method of calculating the parametric function $b_h (T_c, \mu_c)$
which remains still unknown in Eqs. (12-13).
This makes it also possible to calculate the bulk
thermodynamic quantities such as the
critical energy density, $\epsilon_c$, the latent heat, $\epsilon_{LT}$,
etc, in the vicinity of the transition point $T=T_c$.
 
Since the parametric function $b_h (T_c, \mu_c)$ describes details
of the HG phase structure, in order to determine it, one, evidently,
needs to choose the concrete equation of state of the HG phase.
In this part of our work in what follows, $P_h(T_c)$ will denote
the concrete equation of state of the HG phase at $T=T_c$
for point-like particles. In order to take into account extended
volume corrections [18] it should be divided by factor
$[1 + V_0 n_h(T, \mu)]$, where $V_0 = (4 \pi R_0^3 / 3)$ with
$R_0 \sim 0.8 \ fm$, is the volume of nucleon, and $n_h(T, \mu)$
is the baryon number density for the point-like particles, i. e.
it should be calculated from $P_h$ as follows
\begin{equation}
n_h (T, \mu) = \Bigl( {\partial P_h \over \partial \mu} \Bigr)_T.
\end{equation}
The more realistic hadron pressure thus becomes
\begin{equation}
P_H = {P_h \over [1 + V_0 n_h(T, \mu)]}.
\end{equation}
Solving now Eq. (12) against $b_h$ with substitution $P_h
\rightarrow P_H$, one obtains
\begin{equation}
b_h \equiv b_h (T_c, \mu_c) = {(N_f + 1)^2 P_H (T_c, \mu_c) \over
\tilde{B} - (N_f +1) P_H (T_c, \mu_c) }.
\end{equation}
So this relation allows one to calculate $b_h (T_c, \mu_c)$
on account of the chosen equation for the HG phase, $P_H$, by
taking the values of $T_c$ and $\mu_c$ from the general constraint
(11).
Substituting, such calculated value of $b_h (T_c, \mu_c)$, back to
Eqs. (12-13), one obtains the numerical values of $P_q(T_c)$ and
the critical energy density $\epsilon_c$ in our model. In order to
calculate the latent heat, it is necessary first to calculate the
energy density of the HG phase as follows
\begin{equation}
\epsilon_h = T \Bigl( { \partial P_h \over \partial T }
\Bigr)_{\mu}
+ \mu \Bigl( { \partial P_h \over \partial \mu } \Bigr)_T - P_h.
\end{equation}
Similar to $P_H$, the more realistic energy density becomes
\begin{equation}
\epsilon_H = {\epsilon_h \over [1 + V_0 n_h(T, \mu)]}.
\end{equation}
Then the more realistic value of the latent heat is
$\epsilon_{LH} = \epsilon_c - \epsilon_H$,
where $\epsilon_c$ is given by Eq. (13). This is a general scheme
how to calculate $\epsilon_{LH},  \epsilon_c$, etc within the
extended bag model. Let us note that our Ansatz (5-7)
automatically incorporates extended volume corrections.
 
VIII. In this part let us explicitly show how the above
formulated general method works by approximating the hadronic
phase by the ideal (noninteracting) gas of massless mesons.
The state equation in this case is [11, 12, 34]
\begin{equation}
P_h = {1 \over 3} \epsilon_h = g_h {\pi^2 \over 90} T^4,
\end{equation}
where $g_h$ is the number of hadronic degrees of freedom.
Evidently, in this case there are no extended volume corrections,
i. e. $P_h = P_H$ and $\epsilon_h = \epsilon_H$.
From Eq. (29), on account of Eq. (11) at $\mu_c =0$ and Eq. (32)
at $T=T_c$, one finally obtains
\begin{equation}
b_h = { (N_f + 1) g_h \pi^2 \over 30 f_{SB} - g_h \pi^2},
\end{equation}
i. e. in this simple case, the parametric function becomes
constant,
not depending on $T_c$.  The critical energy density (13) then becomes
\begin{equation}
\epsilon_c = {120 f_{SB} - g_h \pi^2 \over 30 (N_f + 1)
        f_{SB}} \tilde{B}.
\end{equation}
The meson gas energy density at $T =T_c$, on account of Eqs. (32)
and (11), is
\begin{equation}
\epsilon_h = 3 P_h =
{ g_h \pi^2 \over 10 (N_f + 1) f_{SB}} \tilde{B},
\end{equation}
and the LH becomes
\begin{equation}
\epsilon_{LH} = \epsilon_c - \epsilon_h =
{30 f_{SB} - g_h \pi^2 \over 7.5 (N_f+1) f_{SB}} \tilde{B}.
\end{equation}
Thus these expressions allow one to calculate the bulk
thermodynamic quantities in the vicinity of the phase
transition in terms of the fundamental quantity, the bag
constant $\tilde{B}$. The numerical results
in the massless pion gas limit are ($g_h = 3$, $N_f= 2$):
$\epsilon_c = 1.358 \ GeV / fm^3, \quad
\epsilon_h = 0.084 \ GeV / fm^3,  \quad
\epsilon_{LH} = 1.274 \ GeV / fm^3$,
in fair agreement with Ref. [34]. This can be explained by the
fact that the authors of the above mentioned paper $a \ priori$
use the value of $T_c$ which coincides
with the value calculated with our method and shown in Eq. (18)
for $N_f=2$. By construction, the phase
transition is of first order, i. e. $\epsilon_{LH}$ is
discontinuous (finite).
 
 Let us now calculate $s_h$ and $c_h$ in this model in order to
estimate the jumps in these quantities between the HG and QGP
phases. The entropy density in the hadronic phase is defined as
$s_h = \partial P_h / \partial T$, where $P_h$ is given by Eq.
(32). On account of the constraint equation (11) at $\mu_c=0$, one
finally obtains
\begin{equation}
s_h (T_c) = {g_h \pi^2 \over 22.5} \Bigl( {3 \tilde{B} \over
             (N_f + 1) f_{SB} } \Bigr)^{3/4}.
\end{equation}
Its numerical value at $g_h =3, N_f=2$ is: $s_h = 0.0054 \ GeV^3$.
The specific heat in the hadronic phase is defined as
$c_h = T \Bigl( \partial^2 P_h / \partial T^2 \Bigr)$. Again on
account of the constraint equation (11) at $\mu_c=0$, one obtains
$c_h(T_c) = 3 s_h(T_c)$, where $s_h(T_c)$ is given by Eq. (37).
So at $g_h =3, N_f=2$  numerically it is $s_h = 0.0162 \ GeV^3$.
In the HG phase these numbers thus are by one order of magnitude
less than the corresponding numbers in the QGP phase (see Eqs.
(26)).
 
 It is instructive to compare the numerical value of $T_c$
which follows from the standard bag model with that of the
extended bag model given in Eqs. (18) for $N_f=2$.
In the standard bag model
the pressure, as given by Eq. (1) at $\mu_f =0$, should
be directly equated to $P_h$, Eq. (32), at $T=T_c$. This gives
the constraint equation as follows
\begin{equation}
T_c = \Bigl( {90 \over 30 f_{SB}- g_h \pi^2} \Bigr)^{1/4}
B^{1/4}.
\end{equation}
Assigning now our value of the bag constant (17) to $B$, for the
massless pion gas limit ($g_h = 3$, $N_f= 2$) one obtains
$T_c \simeq 216 \ MeV$. Of course, this value substantially
contradicts our value (18) as well as the above presented lattice
result (part V).
 
IX. Let us consider now a more sophisticated model of the HG
phase which consists of massless pion gas and nucleons,
antinucleons
with masses $m$. The thermodynamic potential in this case is [10]
\begin{equation}
P_h = {\pi^2 \over 30} T^4 + {4 m^2 \over \pi^2} T^2
\sum_{l=1}^{\infty} {(-1)^{l+1} \over l^2} K_2(l {m \over T})
\cosh(l {\mu \over T}),
\end{equation}
where $K_2$ is the modified Bessel function of the second kind and
the first term describes the massless pion gas limit. This is the
expression for the point-like particles and $P_H$ on account of
the extended volume corrections should be obtained from Eq. (28).
Using now definition (27), one has
\begin{equation}
n_h (T, \mu) = {4 m^2 \over \pi^2} T
\sum_{l=1}^{\infty} {(-1)^{l+1} \over l} K_2(l {m \over T})
\sinh(l {\mu \over T}).
\end{equation}
The energy density $\epsilon_h$, on account of Eqs. (30) and (39),
becomes
\begin{equation}
\epsilon_h = {\pi^2 \over 10} T^4 + {4 m^2 \over \pi^2} T^2
\sum_{l=1}^{\infty} {(-1)^{l+1} \over l^2} [K_2(l {m \over T})
- l {m \over T} K_2 ^{'} (l {m \over T})] \cosh(l {\mu \over T}),
\end{equation}
where prime denotes the differentiation with respect to the
argumentum. Similarly to the previous case, $\epsilon_H$ is to be
obtained from Eq. (31) on account of this expression. These
relations (39-41) are convinient enough to describe the hot HG and
low density matter ($\mu \rightarrow 0$). For cold ($T=0$) and
baryon dense matter, it becomes a fully degenerate Fermi gas at
zero temperature. So the finite temperature corrections can be
found by using a power expansion around $T=0$ [10].
 
Expression (29) for $b_h$, in general, becomes
\begin{equation}
b_h \equiv b_h (T_c, \mu_c) = {(N_f + 1)^2 P_h (T_c, \mu_c) \over
\tilde{B} [1 + V_0 n_h(T_c, \mu_c)]- (N_f +1) P_h (T_c, \mu_c) },
\end{equation}
where $P_h (T_c, \mu_c)$ and $n_h (T_c, \mu_c)$ are given by Eqs.
(39) and (40), respectively with the substitutions $T \rightarrow
T_c$
and $\mu \rightarrow \mu_c$. Thus, in comparison with the previous
case (part VII), now the parametric function is not simply a
constant but it depends crucially on $T_c$ and $\mu_c$ which
should be taken from the general constraint (11) in order to
numerically calculate $b_h(T_c, \mu_c)$ from this expression.
 
  Let us calculate, in this model, $\epsilon_c$ and
$\epsilon_{LH}$
at $\mu_c=0$ which allows us to compare numerical results
with those of the previous part as well as with lattice results
(see below). In this limit extended volume corrections disappear
and one can use classical statistics for nucleons, i. e.
only the first term in the series expansion (39) [10]. On account
of
the constraint equation (11) at $\mu_c=0$, then from Eq. (42), one
finally obtains ($N_f=2$ everywhere below)
\begin{equation}
b_h (T_c) = {3 [{\pi^2 \over 10} + {12 \over \pi^2} m_c^2 K_2(m_c)]
\over f_{SB} (2) -
[{\pi^2 \over 10} + {12 \over \pi^2} m_c^2 K_2(m_c)]},
\end{equation}
where $m_c = m /T_c$ with $m= (m_P + m_N) /2$ and $T_c =160.6 \
MeV$ (see Eqs. (18)). In a similar way, from Eq. (41)
$\epsilon_h$ becomes
\begin{equation}
\epsilon_h (T_c) = {1 \over f_{SB} (2)} \Bigl( {\pi^2 \over 10}
+ {4 m_c^2 \over \pi^2} [ K_2(m_c) - m_c K^{'}_2(m_c)] \Bigl)
\tilde{B}.
\end{equation}
Numerically one get $\epsilon_h = 0.1 \ GeV/fm^3$.
The critical energy density is to be calculated from Eq. (13) on
account of the numerical value of $b_h$, obtained from expression
(43). This is
$\epsilon_c = 1.35 \ GeV/fm^3$,
so the numerical value of the latent heat is
$\epsilon_{LH} = 1.25 \ GeV/fm^3$.
The value of the bag constant as given in Eqs. (17) for
$N_f=2$ was usedi n all numerical calculations above.
These values differ only slightly from those obtained in
the previous
part. Despite the simple models considered above, these numbers
look quite reasonable and they may be compared with lattice result
especially for $\epsilon_c$, discussed for example, in Ref. [35].
The latent heat in this model also remains discontinuous, thereby
confirming the first order nature of the phase transition. The
entropy density $s_h$ as well as specific heat $c_h$, in this
model also differ very slightly from those calculated in the
previous model.
 
 An extension of our model to the case of running coupling
constant
in order to treat interacting QGP (which has been already rather
tentatively discussed in Ref. [36]) along with an extrapolation
outside the phase boundary in order to constract equilibrium phase
transition [10] are subjects of the subsequent papers.
 
  The authors would like to thank J. Zim\'anyi,
K. T\'oth, Gy. Kluge, T. Bir\'o, P. L\'evai, and T. Cs\"org\H o
for useful discussions, remarks and support.

\vfill
\eject

\vfill
\eject

\begin{figure}
 
\caption{The phase diagram in the plane $(T_c, \mu_c)$ measured
         in inits of $MeV$. Here and in all diagrams below the
         curves are only shown for the physically relevant case of
         the two light quarks $N_f=2$ (solid lines). Also shown
         the path of constant specific entropy per baryon $ s/ n_B
         \simeq 50$ (dashed-dotted line).}
 
\bigskip
 
\caption{The entropy density as a function of $T_c$ (upper figure)
         and $\mu_c$ (lower figure).}
 
\bigskip
 
\caption{The specific heat as a function of $T_c$ (upper figure)
         and $\mu_c$ (lower figure). }
 
\bigskip
 
\caption{The baryon number density as a function of $T_c$ (upper
         figure) and $\mu_c$ (lower figure). }

\bigskip
 
\caption{The specific entropy per baryon $ s/ n_B$ as a
         function of $T_c$ (upper
         figure) and $\mu_c$ (lower figure). }
 
\end{figure}

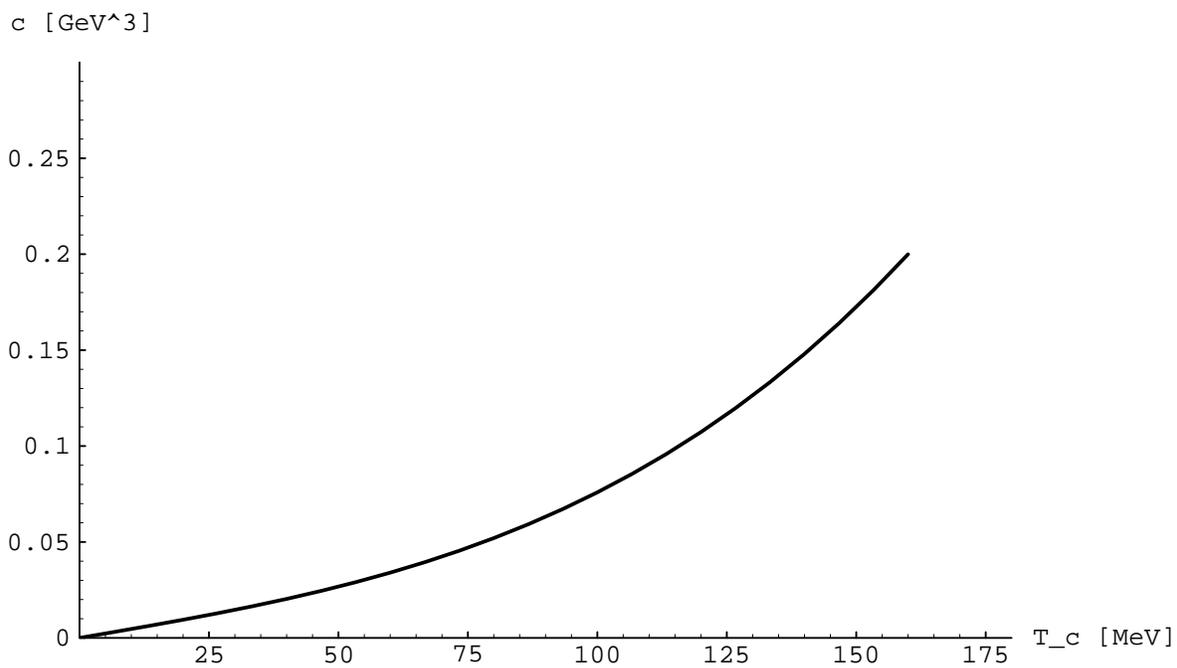
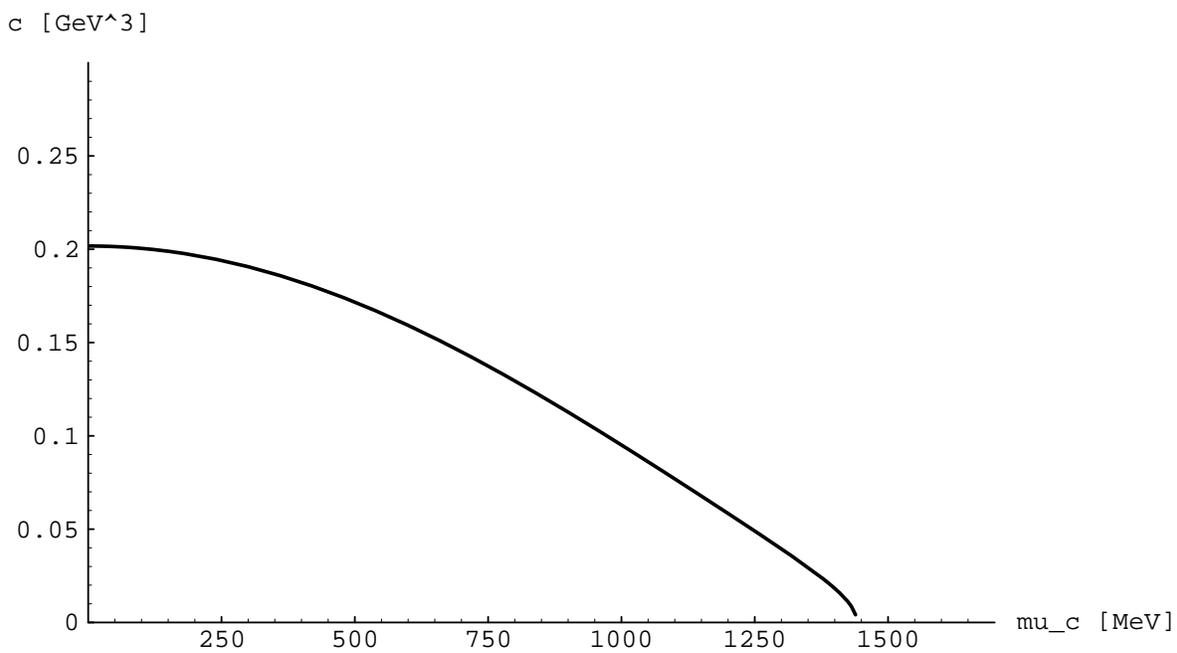
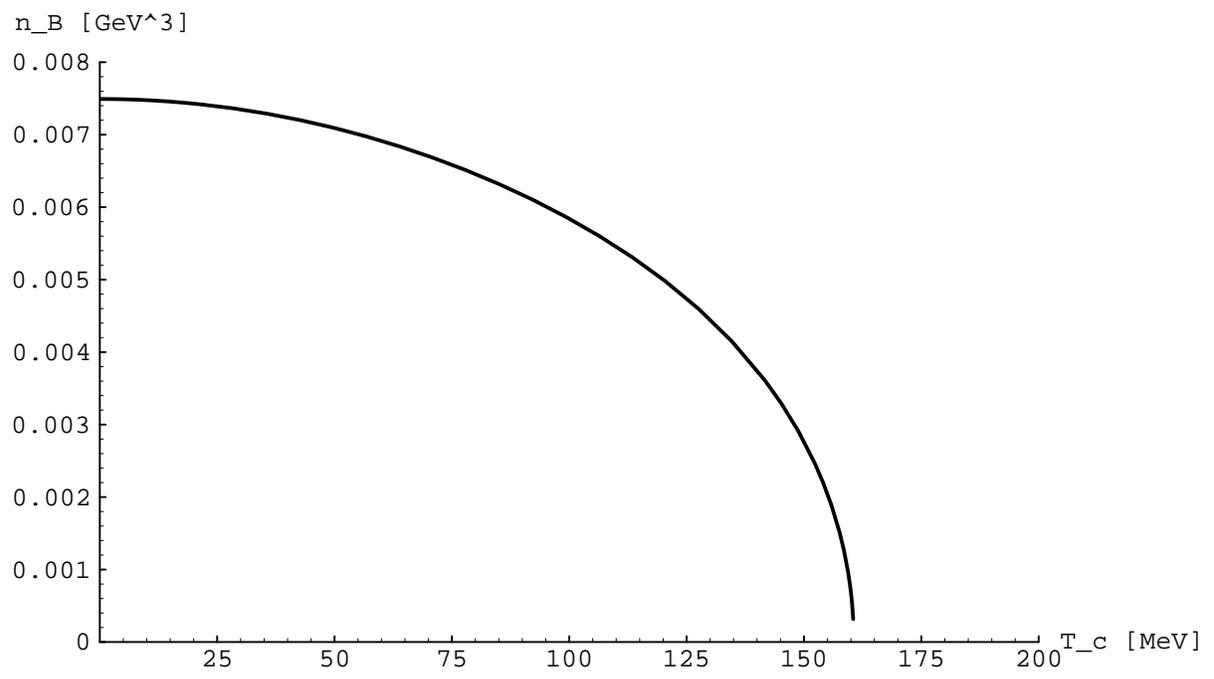
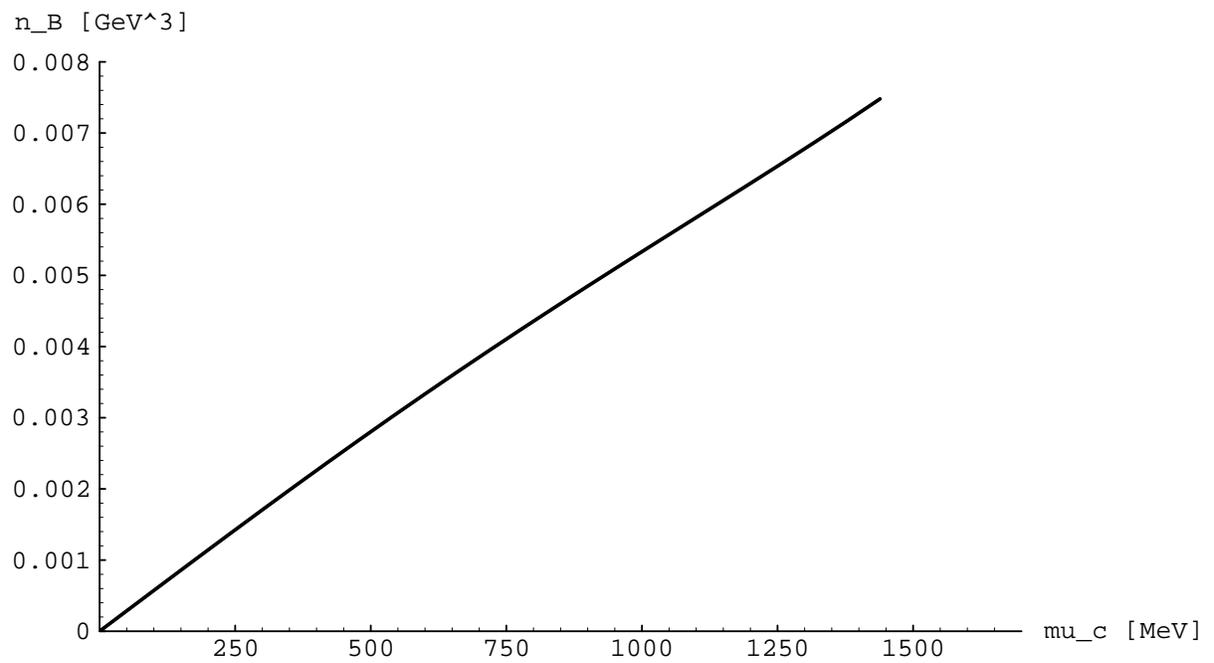
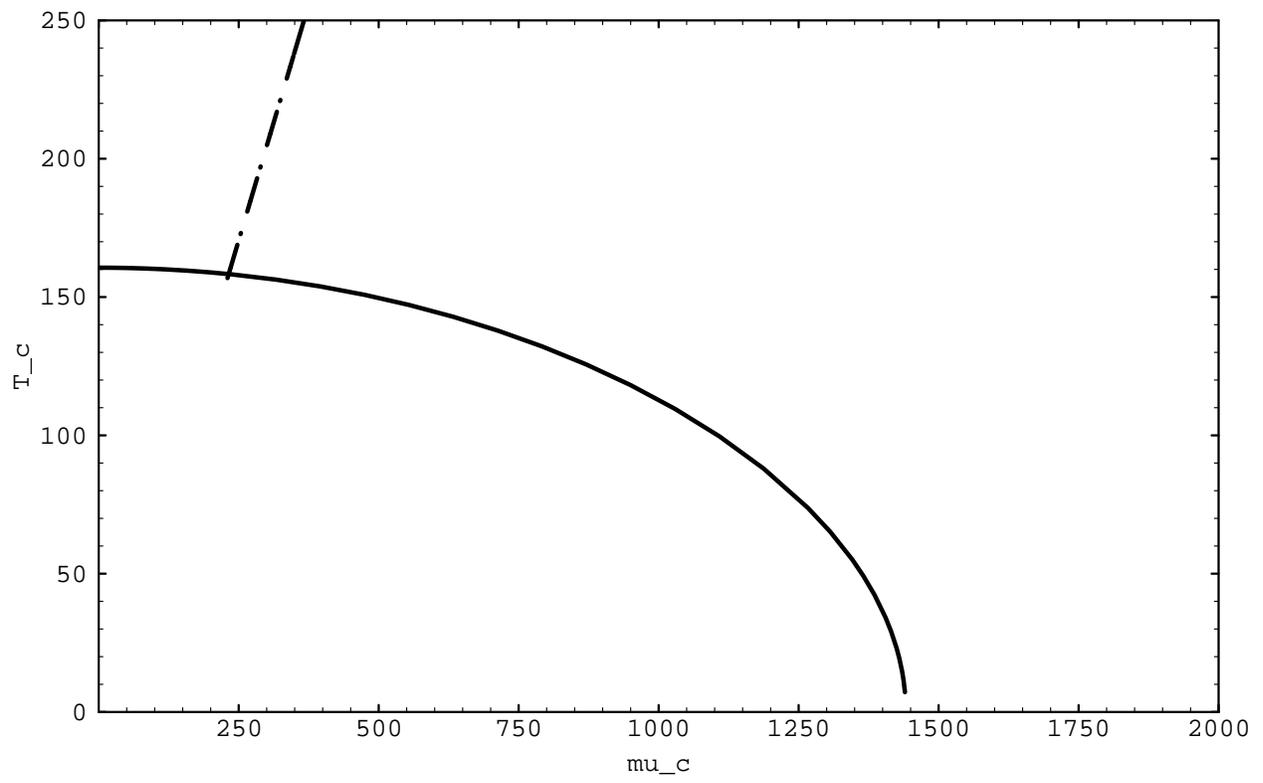
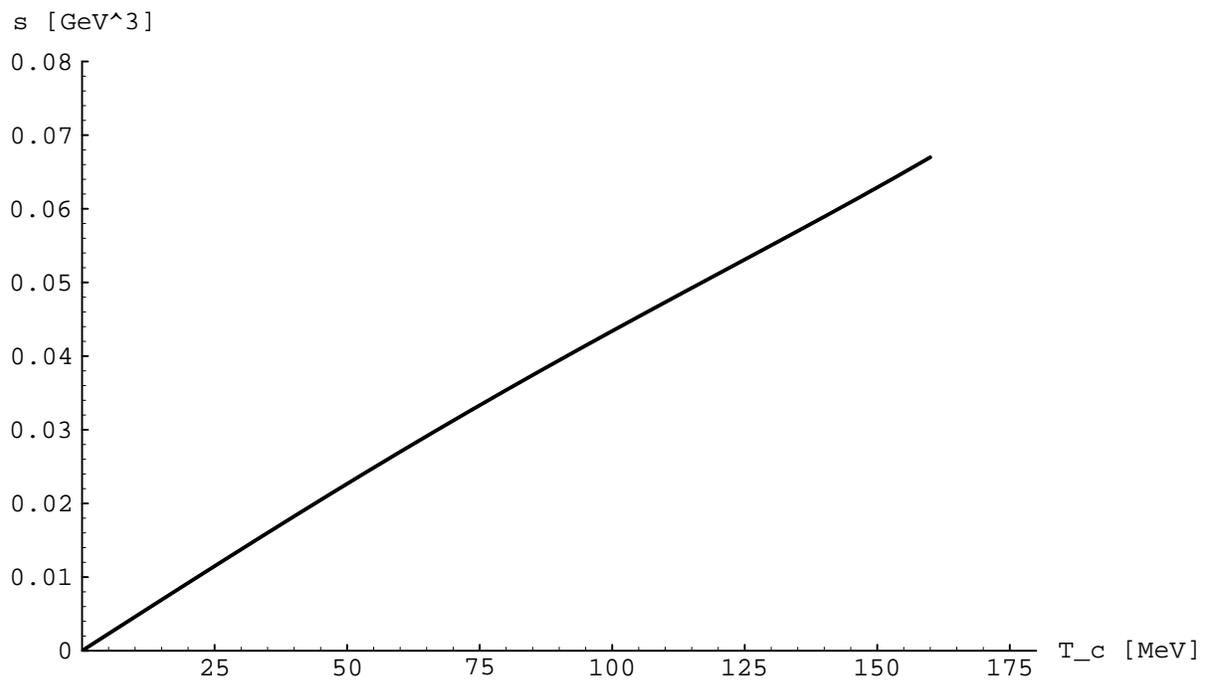
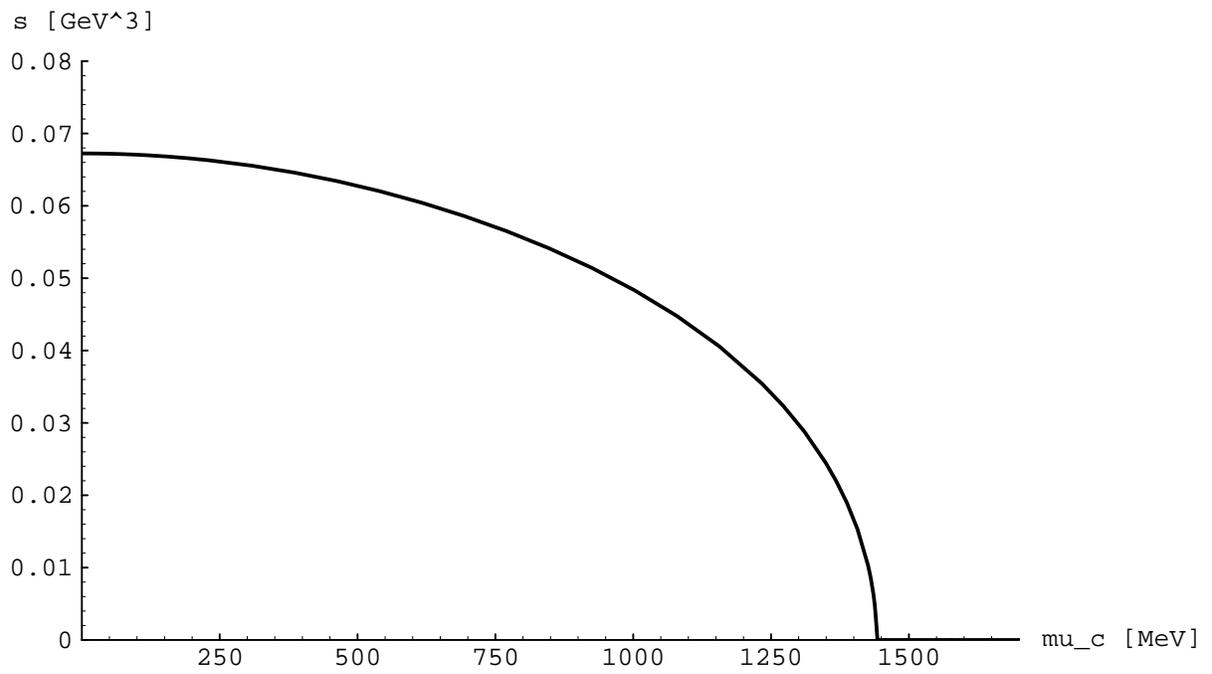
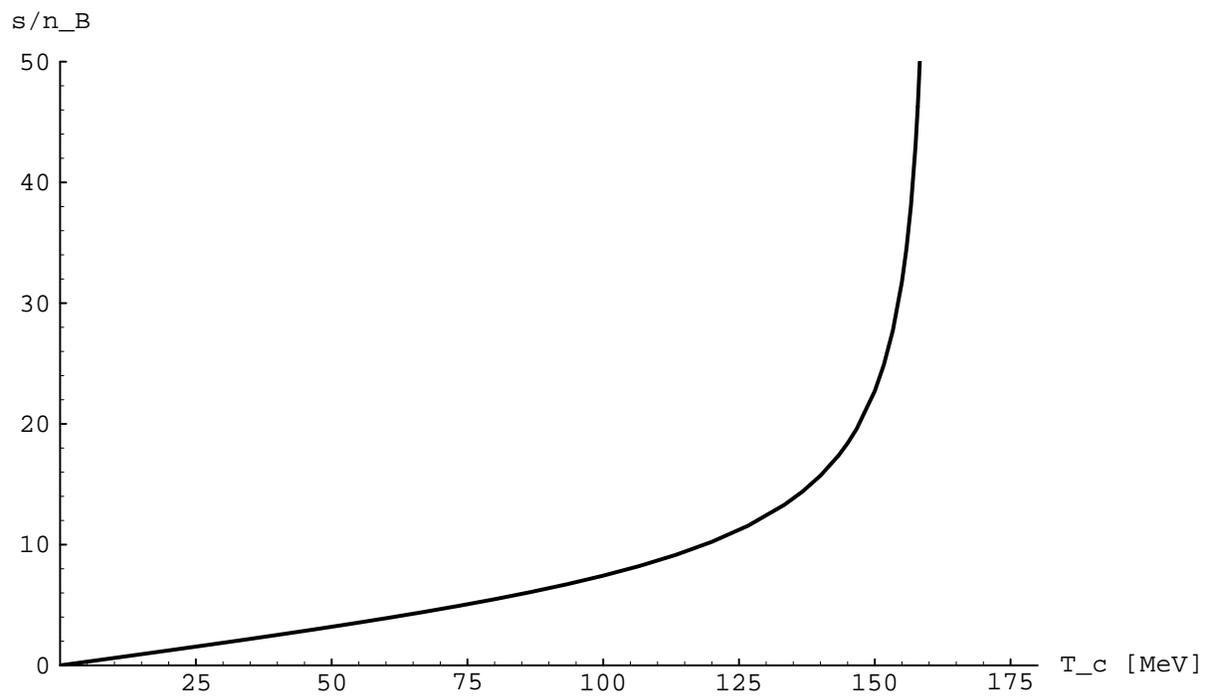
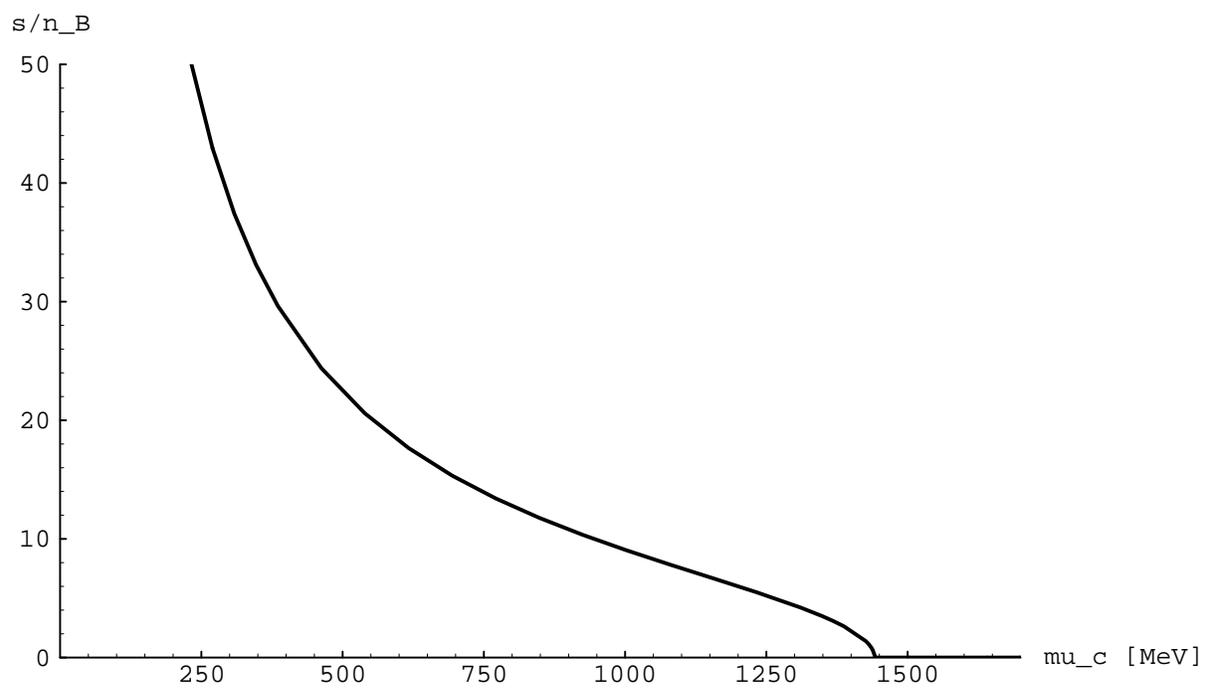
\end{document}